\documentclass{appolb}
\usepackage{epsfig}

\begin{document}
\title{Event by event physics in ALICE%
\thanks{Presented at the "IV Workshop on Particle Correlations and Femtoscopy" 
- Krakow, September 11-14 2008}%
}
\author{Panos Christakoglou
\address{NIKHEF - Utrecht University - FOM}
\and
the ALICE Collaboration.
}
\maketitle
\begin{abstract}
Fluctuations of thermodynamic quantities are fundamental for the study 
of the QGP phase transition. The ALICE experiment is well suited for 
precise event-by-event measurements of various quantities. In this 
article, we review the capabilities of ALICE to study the fluctuations 
of several key observables such as the net charge, the temperature, and 
the particle ratios. Among the observables related to correlations, we 
review the balance functions and the long range correlations.
\end{abstract}
  
\section{Introduction}

Lattice QCD calculations predict that in an environment characterized by 
high temperature and energy density a new state of matter can emerge, where 
the degrees of freedom are given no more by the hadrons but by their 
constituents, the quarks and the gluons (quark gluon plasma - QGP) 
\cite{Ref:QGP}. Experimentally, important information about the nature and 
the time evolution of this predicted phase transition can be extracted by 
studying the final system which emerges after a relativistic heavy ion 
collision. Among other observables, the study of correlations and 
fluctuations on an event-by-event basis is expected to provide additional 
information on the order of this transition \cite{Ref:EbyE}.


\section{Event by event studies in ALICE}

ALICE \cite{Ref:AlicePPR}, located at the CERN LHC, is a multi-purpose 
experiment with highly sensitive detectors around the interaction point. 
The central detectors that cover the region $|\eta| < 0.9$, provide 
good reconstruction and particle identification capabilities as well 
as momentum measurements for each particle species in every event. The 
forward detectors extend the coverage of charged particles and photons. 
A combination of the information given by these detectors provides 
excellent opportunity to study the fluctuations and correlations of 
physics observables on an event-by-event basis at the LHC \cite{Ref:AlicePPR}.


In the next paragraphs, we will review the ongoing studies about some of 
the key event-by-event topics, such as the net charge, and the temperature 
fluctuations, the balance functions, the fluctuations of particle ratios 
and the long range correlations.

\subsection{Fluctuations of the net charge}

Fluctuations of conserved quantities such as the electric charge provide 
information about the initial stage of the formation of the system after 
a collision, when possibly a phase with different degrees of freedom
existed. Both experiments at SPS \cite{Ref:NetChargeNA49} and at RHIC 
\cite{Ref:NetChargeSTAR} have reported that the initial fluctuations can 
be masked by the presence of final state effects (e.g. resonances).

\begin{figure}[ht]
\begin{center}
\epsfig{angle=0,file=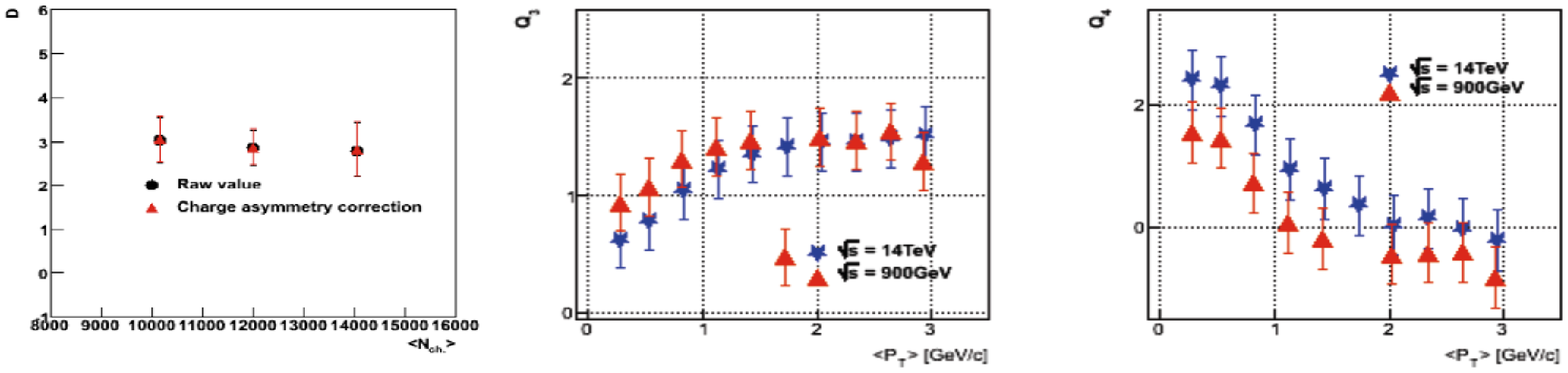,width=12cm,height=4cm}
\end{center}
\caption{The multiplicity dependence of the variable $D$ (left plot) 
calculated for HIJING Pb+Pb collisions at $\sqrt{s_{NN}} = 5.5$ TeV.
The $P_T$ dependence of the higher moments of the net charge distribution 
($Q_3$ middle and $Q_4$ right plot) (PYTHIA p+p collisions at different 
energies).}
\label{Fig:AliceNetChargeFluctuations}
\end{figure}

ALICE plans to study different parameters that are proposed as measures of 
the net charge fluctuations, such as the $D$ parameter \cite{Ref:NetChargeKoch} 
the multiplicity dependence of which for simulated Pb+Pb HIJING collisions 
is shown in Fig. \ref{Fig:AliceNetChargeFluctuations}. In addition, we plan 
to study the higher moments of the net charge (Fig. 
\ref{Fig:AliceNetChargeFluctuations}), such as the skewness and kurtosis 
in order to to explore possible discontinuities in their dependence on 
different parameters \cite{Ref:AlicePPR}.

\subsection{Temperature fluctuations}

The temperature fluctuations provide information about whether there 
is a unique freeze-out temperature of the system or if this parameter 
fluctuates from event to event. The study can be performed either by 
extracting the information from the shape of the $P_T$ distributions 
or by relating the temperature with the event’s transverse mass and 
extracting the variance as proposed in \cite{Ref:TemperatureFluctuations}.

\begin{figure}[ht]
\begin{center}
\epsfig{angle=0,file=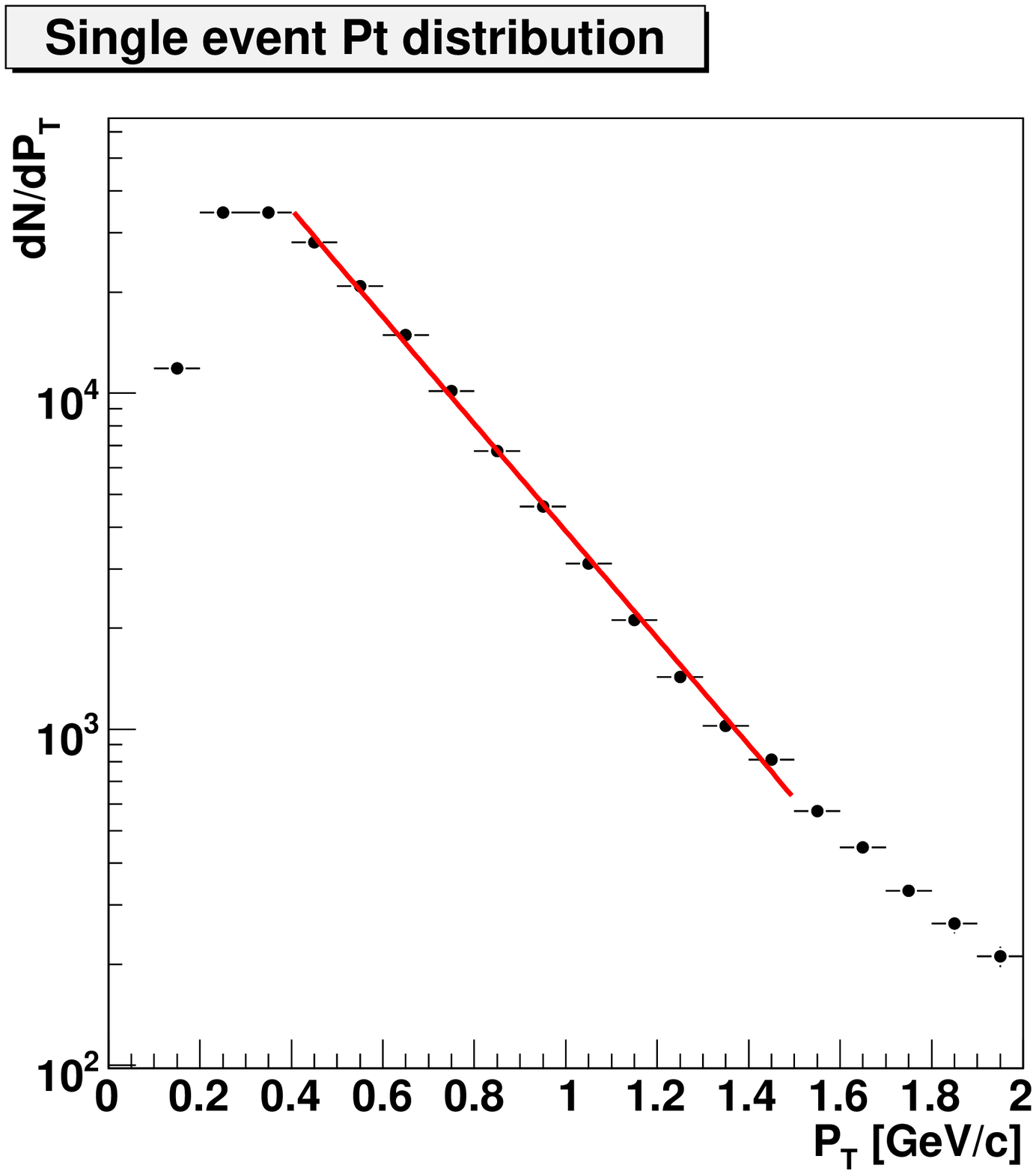,width=4cm,height=4cm}
\epsfig{angle=0,file=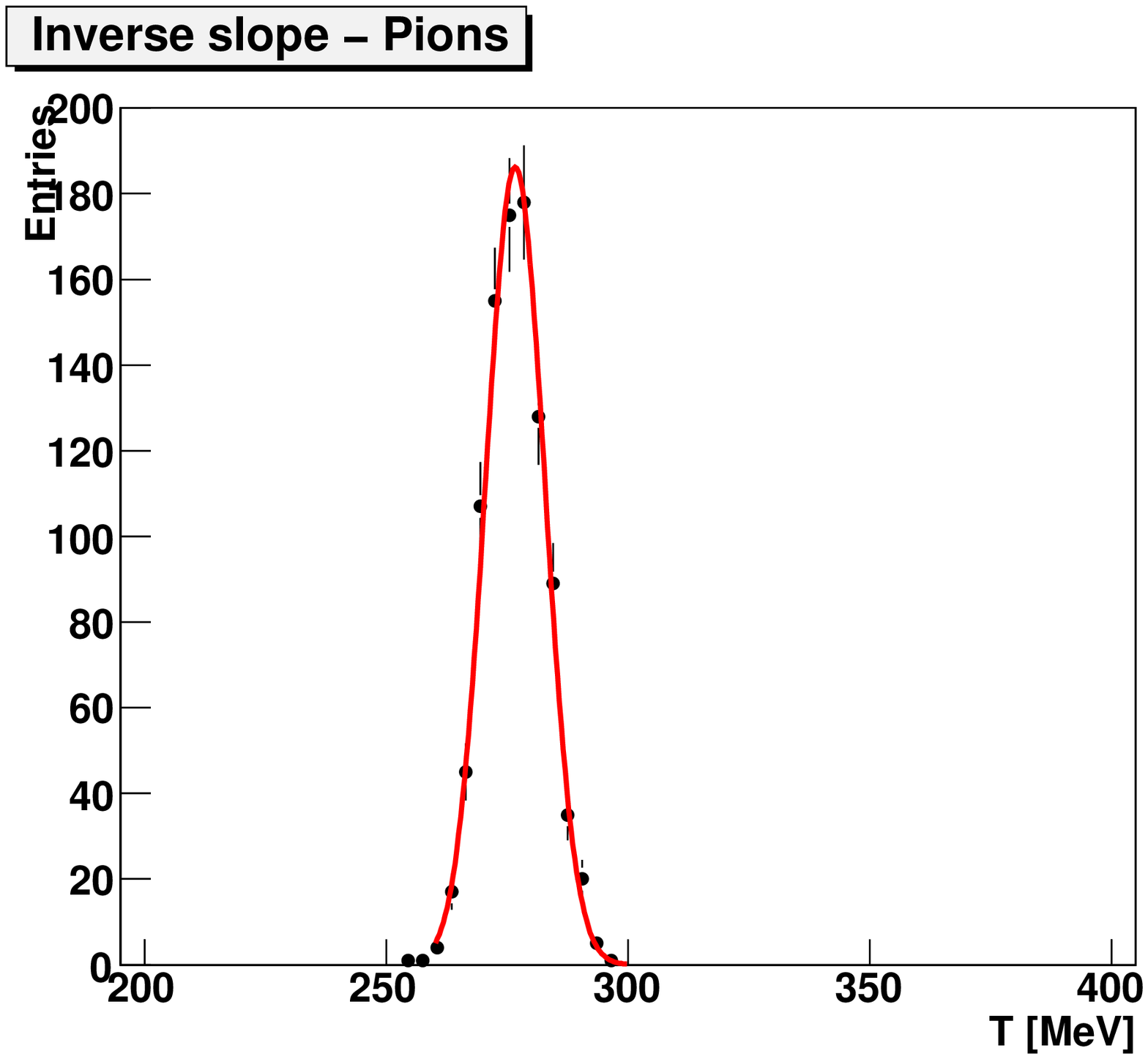,width=4cm,height=4cm}
\epsfig{angle=0,file=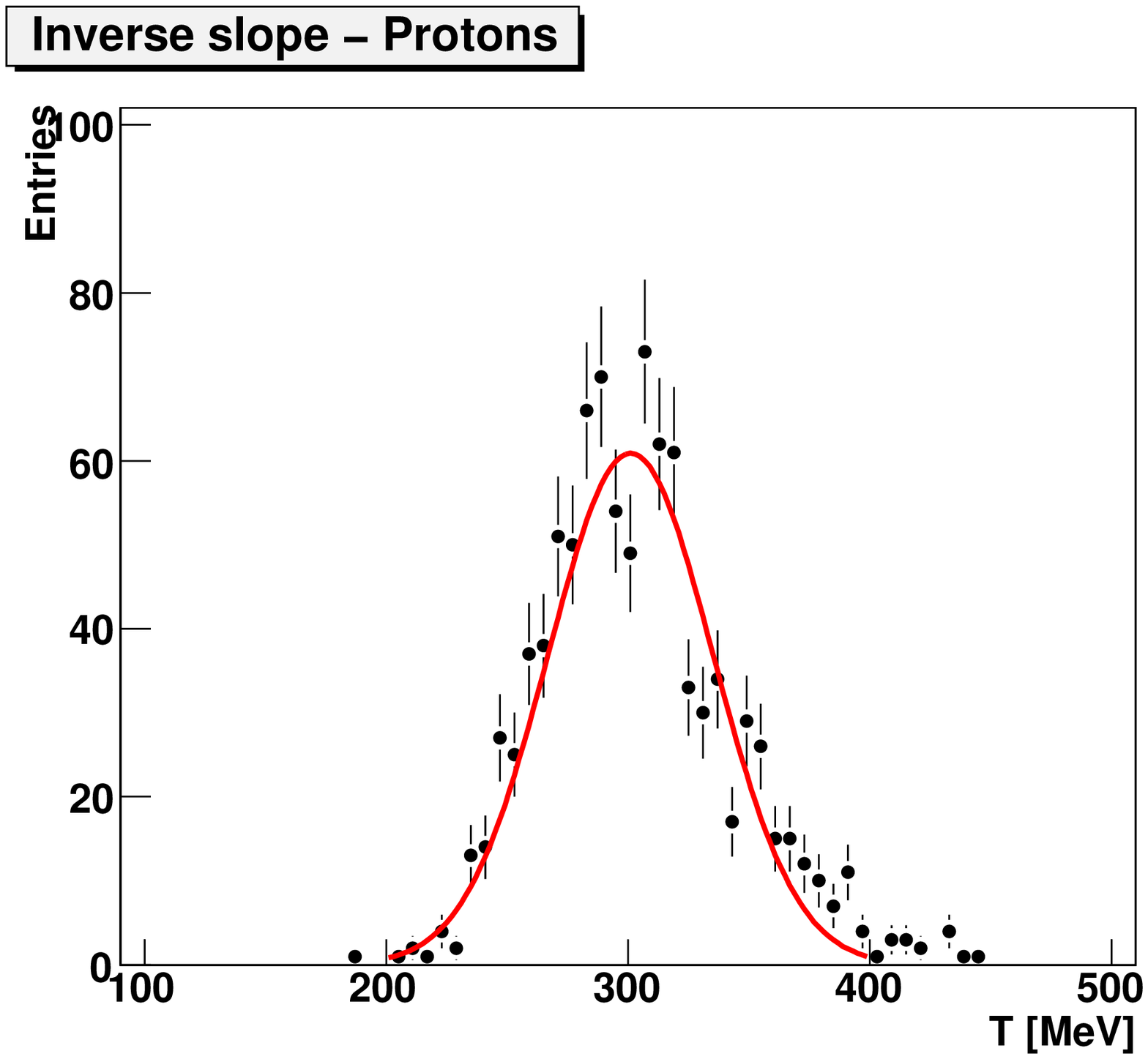,width=4cm,height=4cm}
\end{center}
\caption{The distributions of the inverse slope parameter extracted 
from the single $P_T$ distribution on an event-by-event basis (left plot) 
for identified pions (middle plot) and protons (right plot) (ALICE 
simulations).}
\label{Fig:AliceTemperatureFluctuations}
\end{figure}

Fig. \ref{Fig:AliceTemperatureFluctuations} shows the distributions of 
the inverse slope parameter extracted from an exponential fit to the 
single event $P_{T}$ distribution on an event-by-event basis for pions 
and protons. The results are obtained after the analysis of central 
HIJING Pb+Pb collisions at $\sqrt{s_{NN}} = 5.5$ TeV based on ALICE 
simulations.

\subsection{Balance functions}

The balance function is mainly used to study the event-by-event correlations 
of opposite charged particles but it can also be extended to strangeness or 
baryon number correlations. The width of the balance function can be related 
to the correlation length between the oppositely charged particles as they 
are created at the same location in space-time \cite{Ref:BFPratt}, thus having 
an indirect connection between the width and the time of the hadronization. 
The balance function has been studied by both the NA49 \cite{Ref:BFNA49} and 
the STAR \cite{Ref:BFSTAR} experiments, where both reported a decrease of the 
width with centrality. 

\begin{figure}[ht]
\begin{center}
\epsfig{angle=0,file=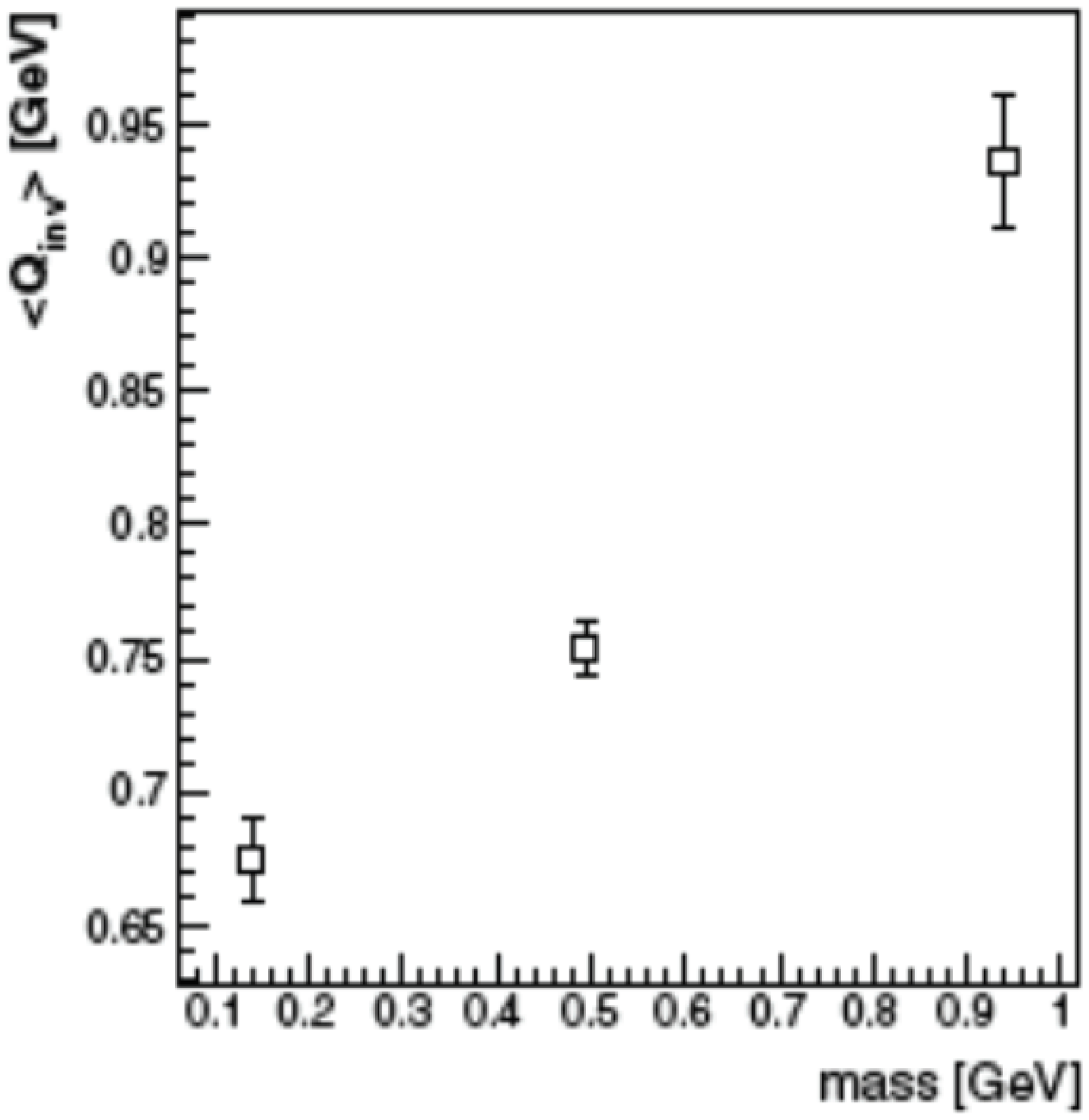,width=4cm,height=4cm}
\epsfig{angle=0,file=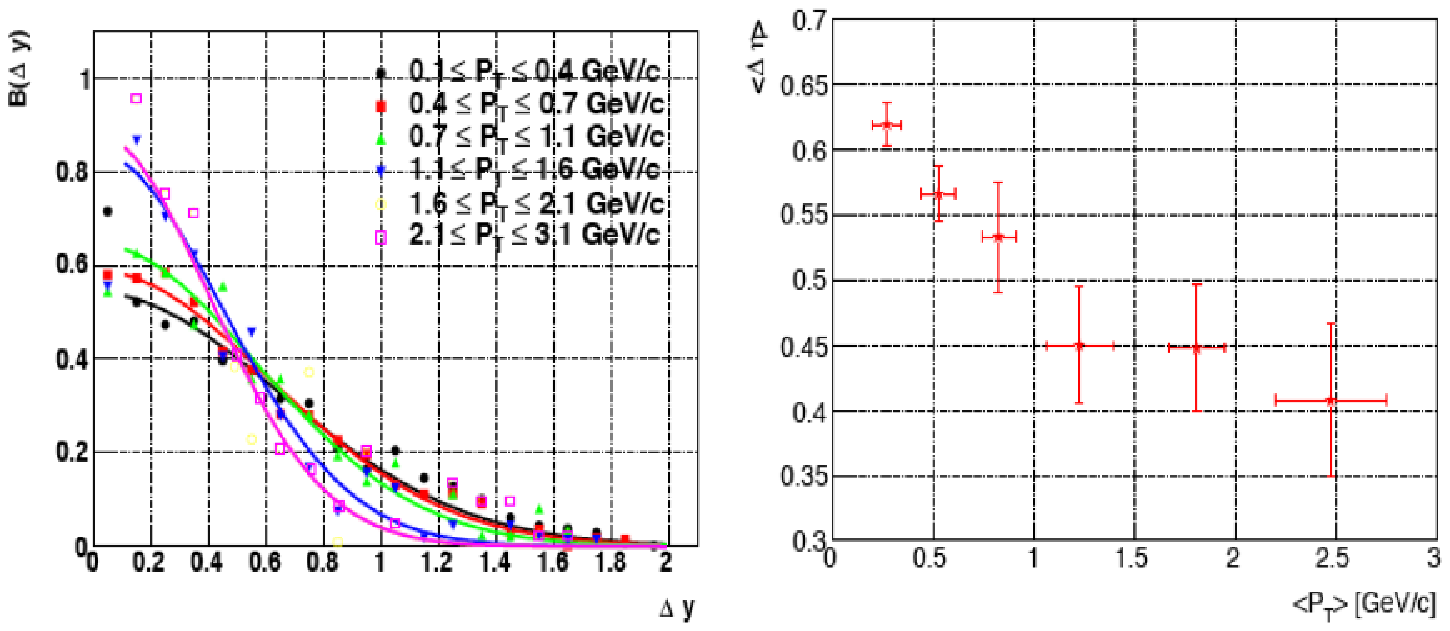,width=8cm,height=4cm}
\end{center}
\caption{The dependence of the $<Q_{inv.}>$ on the mass of the particle that 
form the pairs (left plot). The balance function distributions as a function 
of $\Delta y$ for different $P_T$ intervals (middle plot) and the dependence 
of the width on the $<P_T>$ (ALICE simulations).}
\label{Fig:AliceBalanceFunction}
\end{figure}

For ALICE the balance function will be studied for non-identified charged 
particles but also for different particle species. The balance function will 
also be studied as a function of the $Q_{inv}$ 
(Fig. \ref{Fig:AliceBalanceFunction} left plot) and its components as proposed 
in \cite{Ref:BFQinv}, providing a clearer physics interpretation since these 
parameters tend to be less sensitive to mechanisms that have been proposed as 
being responsible for the narrowing of the width of the balance function 
(e.g. transverse flow). The balance function study can also be extended to a 
two-dimensional differential analysis, in $\eta-P_{T}$ (Fig. 
\ref{Fig:AliceBalanceFunction} middle and right plot) or $\eta-\phi$ space to
 probe the effect of the radial flow \cite{Ref:BFBozek}.

\subsection{Particle ratios}

The particle ratios , especially those that contain information about 
strangeness, are sensitive to the QCD phase transitions. The measure of 
the fluctuations is the $\sigma_{dyn.}$ and is related to the variance of 
the distribution of a given particle ratio for real data and mixed events. 
The ratio of the $K/\pi$ has been studied by different experiments and has 
been reported to decrease with energy throughout the SPS energy range 
\cite{Ref:ParticleRatioNA49} and then stay constant at RHIC energies 
\cite{Ref:ParticleRatioRHIC}.

\begin{figure}[ht]
\begin{center}
\epsfig{angle=0,file=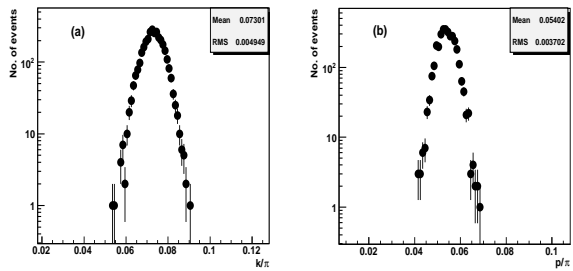,width=8cm,height=4cm}
\end{center}
\caption{The $K/\pi$ and $p/\pi$ ratios calculated event-by-event 
for central HIJING Pb+Pb collisions at $\sqrt{s_{NN}} = 5.5$ TeV 
based on ALICE simulations.}
\label{Fig:AliceParticleRatioFluctuations}
\end{figure}

Fig. \ref{Fig:AliceParticleRatioFluctuations} shows the $K/\pi$ and 
$p/\pi$ ratios calculated event-by-event for central HIJING Pb+Pb 
collisions at $\sqrt{s_{NN}} = 5.5$ TeV based on ALICE simulations. In 
ALICE we can take advantage of the particle identification capabilities 
of the central barrel detectors and extract these ratios by using 
different detector combinations, starting from a stand-alone TPC approach 
and adding detectors reaching the level where the combined PID approach, 
which gives the best results, can be applied.

\subsection{Long range correlations}

The study of correlations among particles produced in different rapidity 
regions can give insight about the particle production mechanism. The 
production of particles in the central region is dominated by short 
range correlations (SRC) at all energies whereas the long range 
correlation (LRC) may be enhanced in hadron-nucleus and nucleus-nucleus 
collision compared to  hadron-hadron collision \cite{Ref:LRC} .

\begin{figure}[ht]
\begin{center}
\epsfig{angle=0,file=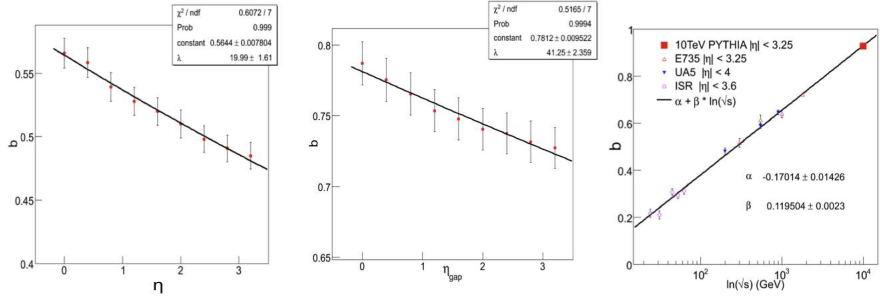,width=12cm,height=4cm}
\end{center}
\caption{The dependence of the parameter b on the pseudo-rapidity keeping 
the $\eta$ gap fixed (left plot) and on the $\eta$ gap keeping the 
pseudo-rapidity interval between the forward and backward regions fixed 
(middle plot). The right plot shows the dependence of the parameter b on 
the energy of the collision (PYTHIA pp events at $\sqrt(s) = 10$ TeV).}
\label{Fig:AliceLongRange}
\end{figure}

ALICE plans to use the central barrel detectors (ITS and TPC) but also 
the forward ones (mainly FMD) for these studies. Fig. \ref{Fig:AliceLongRange} 
shows the dependence of the parameter b, which is defined as the ratio of 
the dispersion of the backward-forward and forward-forward components, on 
the pseudo-rapidity keeping the $\eta$ gap fixed (left plot) and on the 
$\eta$ gap keeping the pseudo-rapidity interval between the forward and 
backward regions fixed (middle plot). The right plot shows the dependence 
of the parameter b on the energy of the collision for real data coming from 
different experiments. These values were fitted with a linear function and 
extrapolated to the LHC energies. The red square corresponds to the result 
obtained from the analysis of PYTHIA pp events at $\sqrt(s) = 10$ TeV.

\section{Conclusions}

The capabilities and prospectives of performing event-by-event measurements 
with the ALICE detectors have been presented motivated also by recent 
lattice QCD calculations predicting that at the very low baryochemical 
potential that corresponds to the LHC energies, interesting fluctuation 
patterns will prevail for heavy ion collisions. ALICE will be very well 
suited to measure the net charge, transverse momentum and temperature 
fluctuations, to study the balance function, to extract the event-by-event 
particle ratios as well as to perform long range correlation studies. In 
addition, the multiplicity the transverse momentum and the azimuthal 
anisotropy fluctuations, the disoriented chiral condensates and the 
fluctuations in the intermediate and high $P_{T}$ sectors will also be 
studied extensively, though not covered in this article.


\end{document}